\documentclass[conference]{IEEEtran}
\IEEEoverridecommandlockouts

\usepackage{cite}
\usepackage{amsmath,amssymb,amsfonts}

\usepackage{graphicx}
\usepackage{textcomp}
\usepackage{xcolor}

\usepackage{multirow}
\usepackage{makecell} 
\usepackage{subfig}

\usepackage{cleveref}
\crefname{figure}{}{}

\usepackage{todonotes}
\usepackage{algorithm}
\usepackage{algpseudocode}

\usepackage{eso-pic}
\newcommand{\paperdoi}{to be assigned (IEEE Xplore)}
\newcommand{\ieeecopyrightnotice}{%
This is a pre-print of the paper accepted for publication in the 2026 17th APCA International Conference on Automatic Control and Soft Computing (CONTROLO). The final published version is available on IEEE Xplore. DOI: \paperdoi}

\def\BibTeX{{\rm B\kern-.05em{\sc i\kern-.025em b}\kern-.08em
    T\kern-.1667em\lower.7ex\hbox{E}\kern-.125emX}}

\begin{document}

\newtheorem{theorem}{Theorem}
\newtheorem{lemma}{Lemma}
\newtheorem{proposition}{Proposition}
\newtheorem{proof}{proof}

\title{Hierarchical Nonlinear Control for Swarm Encirclement Around a Moving Target UAV}

\author{Filipe Cavalheiro$^\dagger$ and Bruno J. Guerreiro$^{\dagger,\ddagger}$
\thanks{$^\dagger$ Filipe Cavalheiro and Bruno Guerreiro are with the Dep. of Electrical and Computer
Eng., CTS-Uninova and LASI, NOVA School of Science and Technology, NOVA University Lisbon, 2829-516
Caparica, Portugal.
E-mails: {\tt\small fs.cavalheiro@campus.fct.unl.pt}, {\tt\small bj.guerreiro@fct.unl.pt}}%
\thanks{$^\ddagger$ B. Guerreiro is also with Institute for Systems and Robotics (ISR), LARSyS, Instituto Superior Técnico, Universidade de Lisboa, Portugal.}%
\thanks{This work acknowledges the financial support of FCT projects CTS(DOI:10.54499/UID/00066/2025 UID/00066/2025), LARSYS (UID/50009/2025), and CAPTURE (DOI:10.54499/PTDC/EEIAUT/1732/2020).}
\thanks{This work acknowledges the use of AI (GPT-5.3 Instant) for code creation and completion being all code verified before use.}
}

\AddToShipoutPictureBG*{\AtPageLowerLeft{\put(0,\LenToUnit{0.35in}){\makebox[\paperwidth]{\parbox{0.84\paperwidth}{\scriptsize\centering\ieeecopyrightnotice}}}}}
\maketitle

\begin{abstract}
This paper presents a hierarchical inner-outer loop nonlinear control strategy using a geometric Lyapunov-based framework for the cooperative encirclement of a target by multiple unmanned aerial vehicles (UAVs) in three-dimensional space. The encirclement problem is formulated on a two-sphere with arbitrary radius and center, denoted $\Gamma^2$, by projecting agent positions from $\mathbb{R}^3$ onto a spherical manifold centered at the target, being the unit two-sphere at the origin $\mathbb{S}^2$ used for intermediate calculations. Desired encirclement configurations are generated via a point separation method on $\Gamma^2$, while stable convergence from $\mathbb{R}^3$ to these points is guaranteed using a Lyapunov-based reduced-attitude control law inspired by geometric control on the two-sphere. Formal stability results are derived using Lyapunov analysis and the framework is validated in simulation using a nonlinear quadrotor model and controller. Results demonstrate stable encirclement and highlight the suitability of the proposed approach for three-dimensional multi-agent aerial systems.
\end{abstract}


\section{Introduction}

Cooperative control of multiple unmanned aerial vehicles (UAVs) has become an important research topic due to applications in surveillance, airspace protection, and interception of non-cooperative aerial agents. Encirclement strategies are attractive because they allow containment and guidance of a target without physical destruction, reducing risks in populated or infrastructure-sensitive environments.

Most existing encirclement and circular formation methods are formulated in two-dimensional spaces, where agents converge to planar circular orbits around a target. Although effective in altitude-constrained scenarios, such approaches do not fully exploit the three-dimensional maneuverability of UAVs. Existing 3D extensions often rely on planar slices or decoupled altitude control, resulting in formations that fail to exploit the full geometry of the surrounding space \cite{jinCircularFormationControl2018,kimCooperativeControlTargetcapturing2007,zhengDistributedControlUniform2015}.

In this work, the encirclement problem is formulated directly on the two-sphere with arbitrary center and radius, denoted $\Gamma^2$, while UAV motion evolves in $\mathbb{R}^3$. Desired configurations are represented as points on a sphere centered at the target, transforming encirclement into a geometric problem on a smooth manifold. This naturally generalizes planar encirclement and enables near-uniform spatial coverage around the target. UAV-to-point assignment is assumed fixed and known, isolating the geometric stabilization problem on $\Gamma^2$. Optimal or decentralized assignment strategies are left for future work. The implementation code is available at \cite{Cavalheiro2026Code}

Encirclement and circular formation control have been widely studied in multi-agent systems. Classical approaches include cyclic pursuit \cite{jinCircularFormationControl2018,kimCooperativeControlTargetcapturing2007,zhengDistributedControlUniform2015}, leader--follower architectures \cite{zhouFixedTimeTargetTracking2026,litimeinSurveyTechniquesCircular2021}, artificial potential fields \cite{litimeinSurveyTechniquesCircular2021}, and virtual structure methods \cite{wangMultiAgentBifurcationConsensusBased2023,litimeinSurveyTechniquesCircular2021}. While these methods provide convergence and spacing guarantees, they are predominantly planar.

Three-dimensional extensions typically decouple radial, angular, and elevation dynamics, yielding formations restricted to planar cross-sections of a sphere \cite{kimCooperativeControlTargetcapturing2007}, or partition the $\mathbb{R}^3$ space into sectors where each UAV independently solves a Tube Model Predictive Control (TMPC) problem \cite{patra3DCooperativePursuit2025}. In parallel, the problem of uniformly distributing points on the two-sphere has been extensively studied in physics and numerical analysis, though these geometric constructions are rarely integrated with nonlinear UAV control laws possessing formal stability guarantees.

Geometric control on manifolds, particularly on the two-sphere, has proven effective for attitude and reduced-attitude stabilization of rigid bodies. Lyapunov-based controllers defined directly on manifolds avoid singularities and provide almost-global stability guarantees. These results motivate the present work, which combines spherical geometry with geometric UAV control for three-dimensional encirclement.

The main contributions of this paper are threefold: (i) integrating point separation methods on $\Gamma^2$ to define encirclement geometries, (ii) adapting geometric Lyapunov-based reduced-attitude controllers to stabilize UAV motion from $\mathbb{R}^3$ onto desired points on $\Gamma^2$, and (iii) validating the approach through simulations using a full nonlinear quadrotor model consistent with existing encirclement literature.

The remainder of the paper is organized as follows. Section~\ref{sec:problem} formalizes the encirclement problem and defines the control objectives. Section~\ref{sec:point-separation} reviews the point separation method used to generate desired configurations on $\Gamma^2$. Section~\ref{sec:controller} presents the proposed controller, Section~\ref{sec:simulations} provides simulation results, and Section~\ref{sec:conclusions} concludes the paper and outlines future research directions.

\section{Problem Formulation} \label{sec:problem}

Let $p_T \in \mathbb{R}^3$ denote the position of a stationary or moving target, and let
$p_i \in \mathbb{R}^3$, $i=1,\dots,N$, denote the position of the $i$-th UAV.

The objective is to drive all UAVs toward desired positions
$p_i^\star \in \mathbb{R}^3$ distributed around the target according to a spherical sampling strategy (Section~\ref{sec:point-separation}), i.e.,
\begin{equation}
\lim_{t \to \infty} p_i(t) = p_i^\star,
\quad \forall i.
\label{eq:control_objective}
\end{equation}

Each desired position lies on a sphere of radius $r^\star > 0$ centered at the target $p_T$,
\begin{equation}
\Gamma^2(p_T,r^\star)
=
\left\{
p \in \mathbb{R}^3
\mid
\|p-p_T\| = r^\star
\right\},
\label{eq:sphere_definition}
\end{equation}
where $\|\cdot\|$ denotes the Euclidean norm.

To separate radial and angular dynamics, define the relative position vector
\begin{equation}
p_i - p_T = d_i = r_i q_i,
\label{eq:relative_decomposition}
\end{equation}
where
$d_i$ is a direction vector, $r_i = \|p_i-p_T\|$ is the distance to the target and
$q_i \in \mathbb{S}^2$ is defined by $q_i = \frac{d_i}{\|d_i\|}$. Similarly, the desired position can be written as
\begin{equation}
p_i^\star = p_T + r^\star q_i^\star,
\label{eq:desired_position_decomposition}
\end{equation}
where $q_i^\star \in \mathbb{S}^2$ denotes the desired angular direction on the two-sphere, being $(\cdot)^\star$ the indicator for desired value.

Therefore, the encirclement objective can be decomposed into the following two sub-objectives:
\begin{equation}
\lim_{t\to\infty} r_i(t) = r^\star,
\qquad
\lim_{t\to\infty} q_i(t) = q_i^\star.
\label{eq:decomposed_objectives}
\end{equation}

Under \eqref{eq:decomposed_objectives}, the original objective
\eqref{eq:control_objective} is recovered directly from
\eqref{eq:desired_position_decomposition}.

\section{Point Separation on the Two-Sphere} \label{sec:point-separation}

Desired encirclement configurations are generated by distributing points on the two-sphere
$\Gamma^2(p_T,r^\star)$. This geometric layer is treated independently from the control design and provides fixed desired positions
$p_i^\star \in \Gamma^2$ for each UAV.

Assuming only the target position is known, the desired configuration should provide approximately uniform coverage around the target, given the uncertainty of motion direction of the target so a well-separated point distribution reduces directional bias and yields similar reachability from all directions.

Among the available spherical sampling methods, the Fibonacci lattice was selected due to its spatial uniformity, flexibility with respect to the number of agents, and low computational cost \cite{gonzalezMeasurementAreasSphere2010}. The method generates quasi-uniform point distributions on the sphere using the golden ratio
\begin{equation}
\varphi = \frac{1+\sqrt{5}}{2}.
\end{equation}

Let $k \in \{0,\dots,P-1\}$ denote the point index, where $P$ is the total number of desired positions. The spherical coordinates of the $k$-th point are defined as
\begin{align}
\mathrm{lat}_k &=
\arcsin\left(
\frac{2k}{P-1}-1
\right), \\
\mathrm{lon}_k &=
2\pi k \varphi^{-1}.
\end{align}

The corresponding Cartesian coordinates on $\Gamma^2(p_T,r^\star)$ are obtained through the standard spherical coordinate transformation.

\begin{figure}[t]
\centering
\subfloat[Normal]{\includegraphics[width=0.32\linewidth]{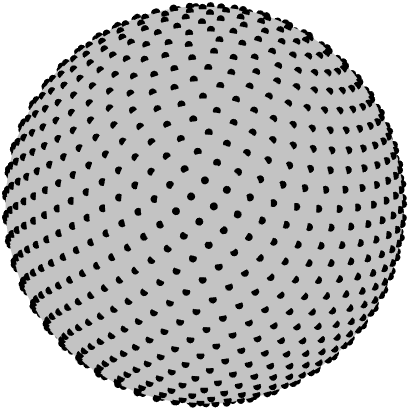}}\hfil
\subfloat[Top]{\includegraphics[width=0.32\linewidth]{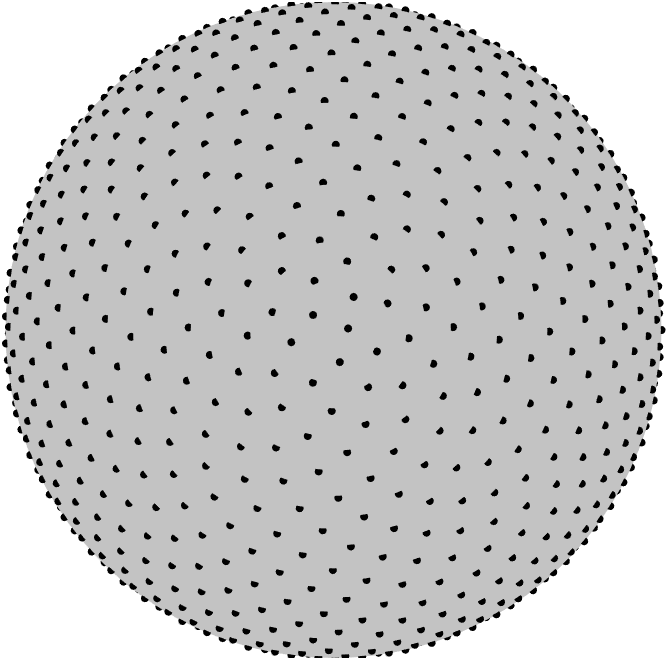}}\hfil
\subfloat[Side]{\includegraphics[width=0.32\linewidth]{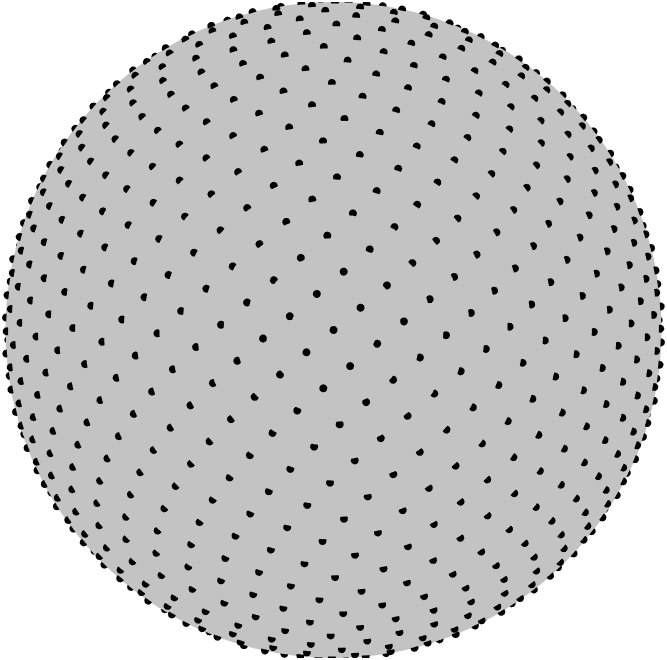}}
\caption{Fibonacci lattice distribution on $\Gamma^2$.}
\label{fig:fib_demo}
\end{figure}

\begin{figure}[t]
\centering
\includegraphics[width=0.85\columnwidth]{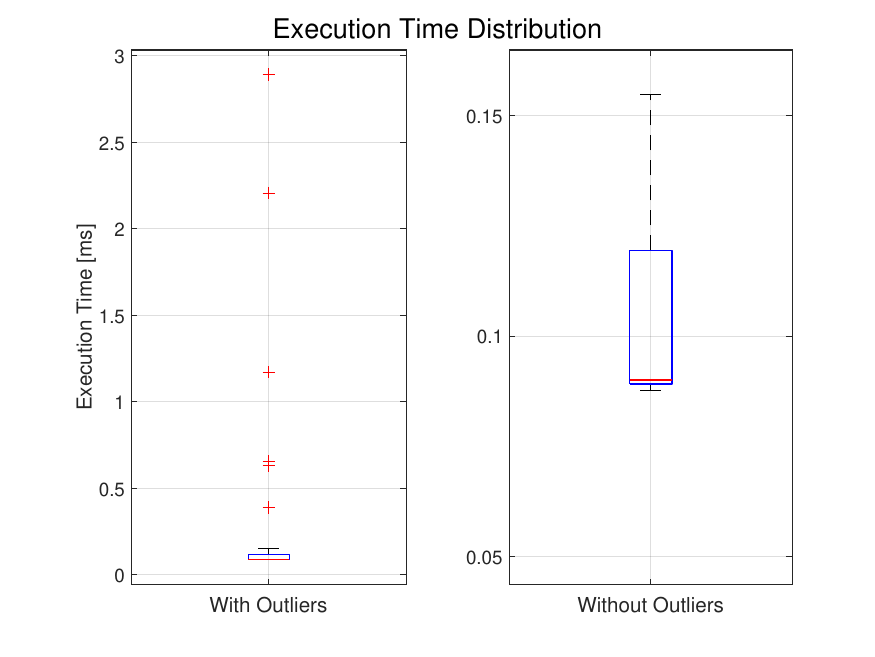}
\caption{Execution time for generating 1001 points over 100 simulations.}
\label{fig:fib_execution_time}
\end{figure}

This construction separates consecutive points by the golden angle, reducing clustering and avoiding periodic alignments. The resulting distribution is highly uniform while remaining computationally inexpensive to generate. An example is shown in Figure~\cref{fig:fib_demo}. Due to its simplicity, good spatial coverage, and negligible computational overhead, as shown in Figure~\cref{fig:fib_execution_time}, the Fibonacci lattice was adopted in this work. The generated points define the desired encirclement geometry, while the proposed control framework, under the assumption of sufficiently accurate inner-loop trajectory tracking, convergence of the guidance dynamics implies convergence of the UAV positions to the desired encirclement configuration.

\section{Controller Design}\label{sec:controller}

The proposed encirclement strategy is implemented using a cascaded inner-outer loop control architecture composed of an inner nonlinear attitude and position drone controller and a outer geometric controller creating a reference for the inner loop moving the drone along the two-sphere.

\subsection{Control Architecture}

\begin{figure}
  \centering
  \includegraphics[width=\columnwidth]{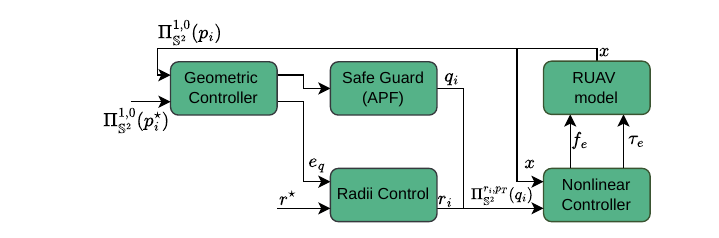}
  \caption{Controller architecture.}
  \label{fig:controller_architecture}
\end{figure}

The overall controller architecture is illustrated in Figure~\ref{fig:controller_architecture}, detailing the information flow between the control blocks. 

The geometric outer-loop controller is responsible for generating high-level motion references that drive each UAV toward its assigned encirclement position $p_i^\star \in \mathbb{R}^3$. These references are defined geometrically on the two-sphere $\Gamma^2$ centered at the target and consist of a desired reduced-attitude direction together with a radial distance profile. The nonlinear (lyapunov based) controller operates purely at the kinematic level and does not depend on the specific UAV dynamics, making it modular and scalable to different aerial platforms.

At each control step, the geometric outer-loop controller computes a desired reference on $\Gamma^2$ with geodesic direction with the artificial potential field (APF) activated when necessary. This reference $q_i$ is passed to the nonlinear controller, which ensures that the UAV model converges to the commanded location while respecting the full nonlinear dynamics.

The previously mentioned APF safeguard is used since the spherical point assignment ensures well-separated desired encirclement locations but transient collisions may still occur during convergence due to arbitrary initial conditions or target motion.

The radial controller is used to reduce the initial distance between the target and each chaser to a user-defined radius. Using the desired radius $r^\star$ and the geodesic error obtained from the geometric controller $e_q$ as inputs, the controller outputs the updated radius $r_i$. Together with the target position and the angular coordinate $q_i$, this updated radius is used to generate the inner-loop reference.

The nonlinear inner-loop controller \cite{mellingerMinimumSnapTrajectory2011} is responsible for tracking the position references generated by the geometric outer-loop controller, being desired velocity set to zero and desired yaw oriented north for all agents. This controller generates the actuation for the Rotary-wing UAV (RUAV) through thrust $f_e$ and body torques $\tau_e$, compensating for nonlinear dynamics, inertial effects, and gravity. The drone state is represented by $x$, which comprises position, attitude, and angular velocity along each axis. However, outside the inner-loop controller, only the position component $p_i$ is used.

\subsection{Coordinate Transformation and Control Design}

This section presents the proposed reduced-attitude control law, which drives each UAV toward its assigned encirclement direction on the two-sphere $\Gamma^2$. The controller is formulated within a geometric Lyapunov framework inspired by the reduced-attitude tracking results in \cite{rampAttitudeAngularVelocity2015}, adopting the same error-function structure and manifold-based formulation.

As in \cite{rampAttitudeAngularVelocity2015}, the objective is to asymptotically regulate the system on $\mathbb{S}^2$ while avoiding the singularities associated with Euler-angle parameterization and the unwinding phenomenon arising in quaternion-based formulations. In the original work, the desired configuration corresponds to a body pointing direction together with a desired angular velocity about that direction. In the present setting, the reference point $p^\star \in \Gamma^2$ represents a stationary target point on the sphere associated with the desired encirclement geometry. Consequently, the desired angular velocity is taken to be identically zero, reducing the problem to a geometric set-point stabilization problem on $\Gamma^2$. Therefore, the resulting controller can be interpreted as a reduced-attitude guidance law that steers the UAV state along geodesic directions on the sphere toward the assigned encirclement point while preserving the global geometric properties of the manifold formulation.

Since spherical projections are used repeatedly throughout the analysis, the operator $\Pi_{\Gamma^2}^{R,P}(\cdot)$ is defined to denote the projection onto a sphere of radius $R$ located at $P$, being for example $\Pi_{\Gamma^2}^{1,0}(p)$ the projection of a point in $p \in \mathbb{R}^3$ onto the unit sphere located at the origin. This operator is formally defined in \eqref{eq:sphere_projections} as $\Pi_{\Gamma^2}^{R,P} : \mathbb{R}^3 \to \Gamma^2(R,P)$ 
\begin{equation}
\Pi_{\Gamma^2}^{R,P}(p)
\;=\;
P + R \frac{p - P}{\|p - P\|},
\qquad p \neq P,
\label{eq:sphere_projections}
\end{equation}
where $\Gamma^2(R,P)$ denotes the two-sphere of radius $R > 0$ centered at $P \in \mathbb{R}^3$, as defined above. Since the Lyapunov controller was designed to work on the unit sphere the relative position vectors encircling the origin with a unit radii is defined as
\begin{equation}
q_i = \Pi_{\Gamma^2}^{1,0}(d_i), \quad q_i^\star =\Pi_{\Gamma^2}^{1,0}(p_i^\star).
\label{eq:hat_q_i}
\end{equation}
The geometric controller objective, previously defined in \eqref{eq:decomposed_objectives} to drive
\begin{equation*}
\lim_{t \to \infty} q_i(t) = q_i^\star,
\end{equation*}
where the motion of each $q_i$ is assumed to be given by
\begin{equation}
\dot{q}_i = -k_{geo} S(e_{q_i}) q_i,
\label{eq:geometric_control}
\end{equation}
where $S(\cdot)$ is the cross-product skew-symmetric matrix satisfying $S(a) b = a \times b$, for $a,b\in\mathbb{R}^3$, $k_{geo} > 0$ is a constant scaler gain and the error $e_{q_i}$ is  defined in \eqref{eq:e_q}, being the error functions introduced by quantifying the reduced-attitude error, consider the smooth configuration error function 
\begin{equation}
\Psi_i(q_i, q_i^\star)
= 2 - 2 \sqrt{\frac{1 + q_i^\top q_i^\star}{2}},
\label{eq:psi}
\end{equation}
as defined in \cite{rampAttitudeAngularVelocity2015}.
While other simpler error functions could be considered, it is noted that the function $\Psi_i$ is positive definite about $\hat{q}_i = \hat{q}_i^\star$, vanishes at the desired configuration and is locally quadratic. The associated configuration error vector, defined on the tangent space, is
\begin{equation}
e_{q_i}
=
\frac{S(q_i^\star) q_i}{\sqrt{2(1 + q_i^\top q_i^\star)}},
\label{eq:e_q}
\end{equation}
which satisfies $e_{q_i} = 0$ if and only if $q_i = q_i^\star$.

The desired position reference for the nonlinear controller is then obtained by projecting $q_i$ back to $\Gamma^2$ as
\begin{equation*}
    d_i = \Pi_{\Gamma^2}^{r_i,p_T}(q_i),
\end{equation*}
which is passed to the safeguard to avoid collisions occurrences, as described in the next section.

\begin{proposition}
Given the error function \eqref{eq:e_q} the equilibrium $q_i = q_i^\star$ of the closed-loop system \eqref{eq:geometric_control} is almost globally asymptotically stable on $\mathbb{S}^2$.
\end{proposition}

\begin{IEEEproof}
Consider the Lyapunov candidate
\begin{equation*}
V(q_i)
=
\Psi(q_i,q_i^\star),
\end{equation*}
which is positive definite about $q_i = q_i^\star$, satisfying
\begin{equation*}
V(q_i) \ge 0,
\qquad
V( q_i) = 0
\iff
q_i = q_i^\star.
\end{equation*}

The error function \eqref{eq:e_q} is well defined for all
$q_i \in \mathbb{S}^2 \setminus \{- q_i^\star\}$, excluding the
antipodal singularity inherent to almost-global stabilization on the sphere.

Consider the standard kinematics on $\mathbb{S}^2$
\begin{equation*}
\dot{q}_i = S(\omega_i) q_i.
\end{equation*}
Comparing with the closed-loop dynamics \eqref{eq:geometric_control} gives
\begin{equation*}
\omega_i = -k_{geo} e_{q_i}.
\end{equation*}

From the geometric properties of the configuration error function,
its time derivative satisfies
\begin{equation*}
\dot \Psi = e_{q_i}^\top \omega_i.
\end{equation*}
Therefore,
\begin{equation*}
\dot V
=
\dot \Psi
=
e_{q_i}^\top(-k_{geo}e_{q_i})
=
-k_{geo}\|e_{q_i}\|^2
\le 0 .
\end{equation*}

Hence, $V$ is non-increasing along the trajectories of the system, and all trajectories remain bounded on the compact manifold $\mathbb{S}^2$. The set where $\dot V = 0$ is characterized by $e_{q_i}=0$.
From \eqref{eq:e_q}, this occurs at
\begin{equation*}
q_i = q_i^\star
\qquad \text{or} \qquad
q_i = - q_i^\star.
\end{equation*}

By LaSalle's invariance principle, every trajectory converges to the largest invariant set contained in $\{q_i \in \mathbb{S}^2 : e_{q_i}=0 \}$. The equilibrium $q_i = q_i^\star$ corresponds to the global minimum of the configuration error function and is asymptotically stable. The antipodal equilibrium $ q_i = -q_i^\star$ corresponds to the maximum of the configuration error function $\Psi$. Therefore, arbitrarily small perturbations away from $q_i^\star$ decrease the Lyapunov function and drive trajectories away from the antipodal configuration under the closed-loop dynamics. Hence, the antipodal equilibrium is unstable. Leading to all trajectories on $\mathbb{S}^2$, except those initialized at the unstable antipodal equilibrium, converge asymptotically to $q_i^\star$. Hence, the equilibrium $q_i = q_i^\star$ is almost globally asymptotically stable on $\mathbb{S}^2$.
\end{IEEEproof}

\subsection{Artificial Potential Field Safeguard}

The safeguard operates directly on the two-sphere $\mathbb{S}^2$ and modifies only the reference position ${q}_i$. This design choice ensures that the safeguard acts as a bounded perturbation to the nominal reduced-attitude reference. Specifically, at each control step, all UAV positions are projected onto the unit sphere centered at the origin
\begin{equation*}
    q_i = \Pi_{\Gamma^2}^{1,0}(d_i), \forall i.
\end{equation*}

Pairwise geodesic distances on $\mathbb{S}^2$ are then computed. Let $dist_{ij}$ denote the spherical geodesic distance between UAV $i$ and all other UAVs $j$. A minimum safe separation $dist_{\min} > 0$ is defined, in this case half of the the average geodesic distance between neighboring Fibonacci points $p^\star$ is used. For each neighboring UAV satisfying $dist_{ij} < dist_{\min}$, a repulsive tangential displacement is generated. Let
\begin{equation*}
    n_i = \frac{q_i - p_T}{\|q_i - p_T\|},
\end{equation*}
denote the outward normal direction at UAV $i$.
The tangential component of the relative displacement vector is defined as
\begin{equation*}
    \Delta_{ij}^{tan}
    =
    (q_{i} - q_{j})
    -
    ((q_{i} - q_{j})^{\top} n_i) n_i,
\end{equation*}
ensuring that the safeguard does not interfere with radial stabilization. The repulsive contribution from agent $j$ to agent $i$ is defined as
\begin{equation*}
    f_{ij}
    =
    (dist_{\min} - dist_{ij})
    \frac{\Delta_{ij}^{tan}}{\|\Delta_{ij}^{tan}\|},
    \quad dist_{ij} < dist_{\min},
\end{equation*}
being the total artificial potential displacement obtained by summing all active pairwise contributions
\begin{equation*}
    f_{total} = \sum_{j \neq i} f_{ij}.
\end{equation*}

To prevent excessive deviation from the nominal reference, the resulting displacement is saturated to a bounded magnitude of a maximum of $d_{min}$ and the corrected reference is re-projected onto the sphere:
\begin{equation}
    d_i =
    \Pi_{\Gamma^2}^{r_i,p_T}
    \left(
        q_{i} + f_{total}
    \right).
    \label{eq:apf_result}
\end{equation}

\begin{figure}
    \centering
    \subfloat[Active]{\includegraphics[width=0.42\columnwidth]{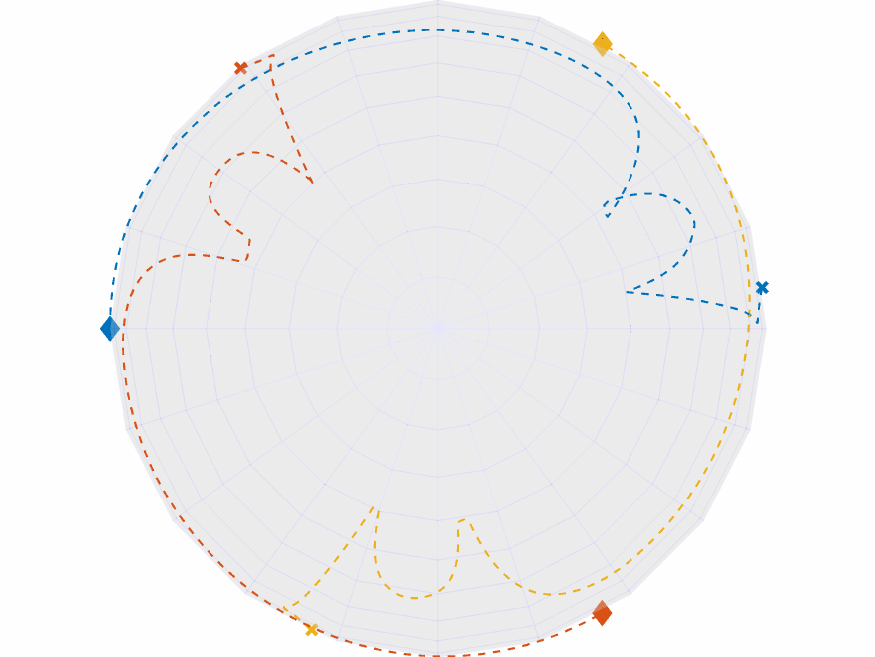}}
    \subfloat[Inactive]{\includegraphics[width=0.42\columnwidth]{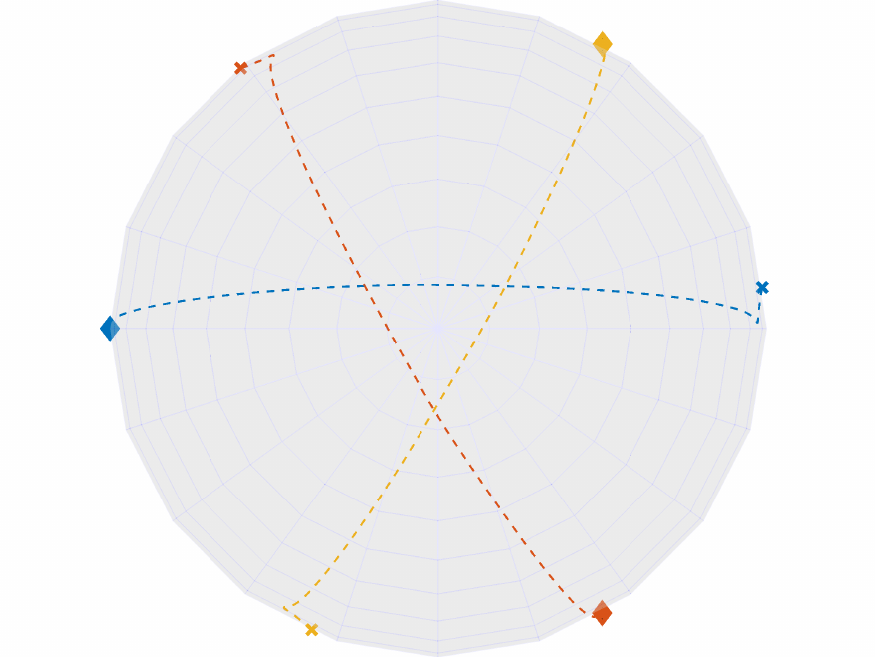}}
    \caption{Demonstration of APF collision avoidance given near collision trajectories.}
    \label{fig:AFP_test}
\end{figure}

The APF safeguard acts as a bounded transient perturbation designed to improve collision avoidance during convergence, producing a modified reference direction
$d_i$, since the APF contribution is bounded and vanishes outside the collision region, the nominal asymptotic behavior is recovered once inter-agent distances exceed the safety threshold. In practice, this mechanism, as shown in Figure~\ref{fig:AFP_test}, alleviates collision avoidance during convergence while allowing the Lyapunov-based controller to recover the desired encirclement configuration once agents are safely separated.

\subsection{Radii control}

As seen in \eqref{eq:decomposed_objectives} the objective for the controller was separated in two parts having the motion on the two sphere updated the radial update happens as  define by the adaptive law \eqref{eq:radial_control}
 
\begin{equation}
\dot{r}_i = k_{\mathrm{rad}} e^{-( \|e_{q_i}\|)^2}
\left( r^\star - r_i \right),
\label{eq:radial_control}
\end{equation}
being $e_{q_i}$ the reduced-attitude error defined in \eqref{eq:e_q} and defining $k_{\mathrm{rad}} > 0$ as constant scaler gain.

\begin{proposition}
Consider the closed-loop system composed of the reduced-attitude dynamics \eqref{eq:geometric_control} and the radial dynamics \eqref{eq:radial_control} for all $i = 1,\dots,N$, where $k_{\mathrm{geo}} > 0$, $k_{\mathrm{rad}} > 0$, $r^\star > 0$ is constant, and $e^{-\|e_{q_i}\|^2} > 0$. Then, the equilibrium defined by
\begin{equation}
\hat q_i = \hat q_i^\star, \qquad r_i = r^\star,
\end{equation}
is almost globally asymptotically stable.
\label{proposition:combined_stability}
\end{proposition}

\begin{IEEEproof}

Define the radial tracking error
\begin{equation}
e_{r_i} = r_i - r^\star.
\label{eq:radial_err}
\end{equation}

Since $r^\star$ is constant, its derivative satisfies
\begin{equation*}
\dot e_{r_i} = \dot r_i.
\end{equation*}

Substituting the radial control law \eqref{eq:radial_control} gives
\begin{equation*}
\dot e_{r_i}
=
k_{\mathrm{rad}}
e^{-\|e_{q_i}\|^2}
(r^\star-r_i)
=
-k_{\mathrm{rad}}
e^{-\|e_{q_i}\|^2}
e_{r_i}.
\end{equation*}

Using the configuration error function $\Psi_i$ defined in \eqref{eq:psi}, consider the composite Lyapunov candidate
\begin{equation}
V
=
\sum_i
\left(
\Psi_i
+
\frac12 e_{r_i}^2
\right).
\label{eq:composite_lyapunov}
\end{equation}

From Proposition 1, $\Psi_i$ is positive definite about $q_i=q_i^\star$, except at the antipodal configuration $q_i=-q_i^\star$, which is unstable. Furthermore, the quadratic term $\frac12 e_{r_i}^2$ is positive definite with respect to the radial error. Hence, $V$ is positive definite with respect to the desired equilibrium.

Taking the time derivative of $V$ along the closed-loop trajectories yields
\begin{equation*}
\dot V
=
\sum_i
\left(
\dot\Psi_i
+
e_{r_i}\dot e_{r_i}
\right).
\end{equation*}

Using the reduced-attitude result from Proposition 1,
\begin{equation*}
\dot\Psi_i
=
-k_{\mathrm{geo}}\|e_{q_i}\|^2,
\end{equation*}
and the radial error dynamics,
\begin{equation*}
e_{r_i}\dot e_{r_i}
=
-k_{\mathrm{rad}}
e^{-\|e_{q_i}\|^2}
e_{r_i}^2,
\end{equation*}
we obtain
\begin{equation*}
\dot V
=
-\sum_i
k_{\mathrm{geo}}\|e_{q_i}\|^2
-
\sum_i
k_{\mathrm{rad}}
e^{-\|e_{q_i}\|^2}
e_{r_i}^2
\le 0.
\end{equation*}

Since $k_{\mathrm{geo}}>0$, $k_{\mathrm{rad}}>0$, and
$e^{-\|e_{q_i}\|^2}>0$ for all $e_{q_i}$, the derivative $\dot V$ is negative semidefinite and vanishes only when
\begin{equation*}
e_{q_i}=0, \quad e_{r_i}=0, \quad \forall i.
\end{equation*}

The condition $e_{r_i}=0$ implies
\begin{equation*}
r_i=r^\star,
\end{equation*}
while $e_{q_i}=0$ corresponds either to the desired equilibrium
\begin{equation*}
q_i=q_i^\star,
\end{equation*}
or to the antipodal equilibrium
\begin{equation*}
q_i=-q_i^\star.
\end{equation*}

By Proposition 1, the antipodal equilibrium is unstable and therefore excluded from the region of attraction. Applying LaSalle's invariance principle on the manifold $\mathbb{S}^2 \times \mathbb{R}_{>0}$, all trajectories except those initialized at the antipodal set converge asymptotically to
\begin{equation*}
q_i=q_i^\star,
\qquad
r_i=r^\star.
\end{equation*}

Hence, the overall closed-loop system is almost globally asymptotically stable.
\end{IEEEproof}

Given Proposition 2 and the decomposition of the desired position in \eqref{eq:desired_position_decomposition}, together with the decomposed objectives in \eqref{eq:decomposed_objectives}, it follows that, in the absence of actuation from the APF safeguard, the closed-loop system evolves almost globally asymptotically toward the main objective $p_i^\star$.

\section{Simulation Results}\label{sec:simulations}

Two encirclement scenarios were evaluated in simulation: one with 3 drones for ease of visualization and another with 31 drones to demonstrate scalability to a more realistic swarm size. In both cases, the target moves along a half-circle trajectory before becoming stationary at the end of the path.

\begin{figure}
    \includegraphics[width=0.45\columnwidth]{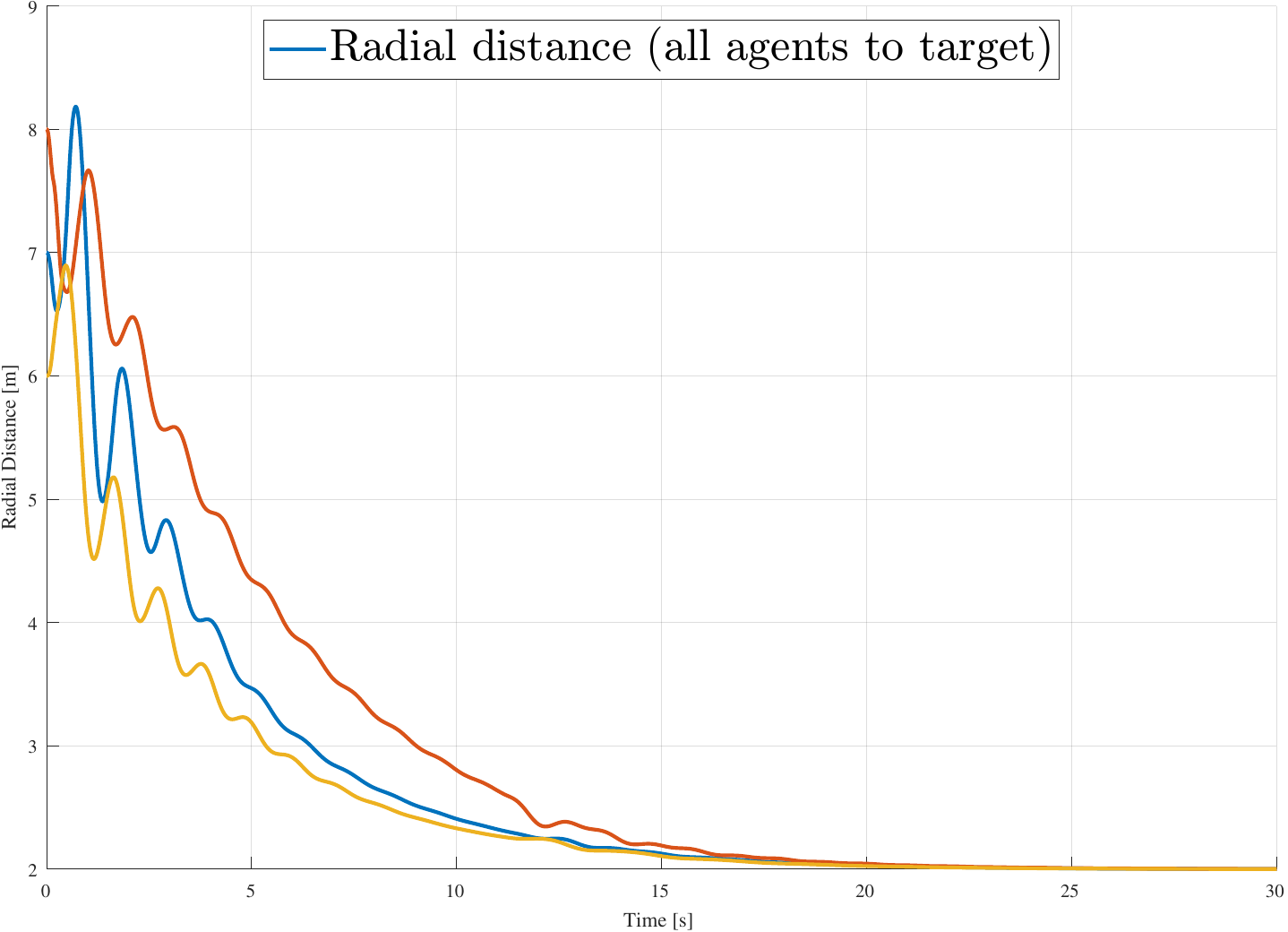}
    \includegraphics[width=0.45\columnwidth]{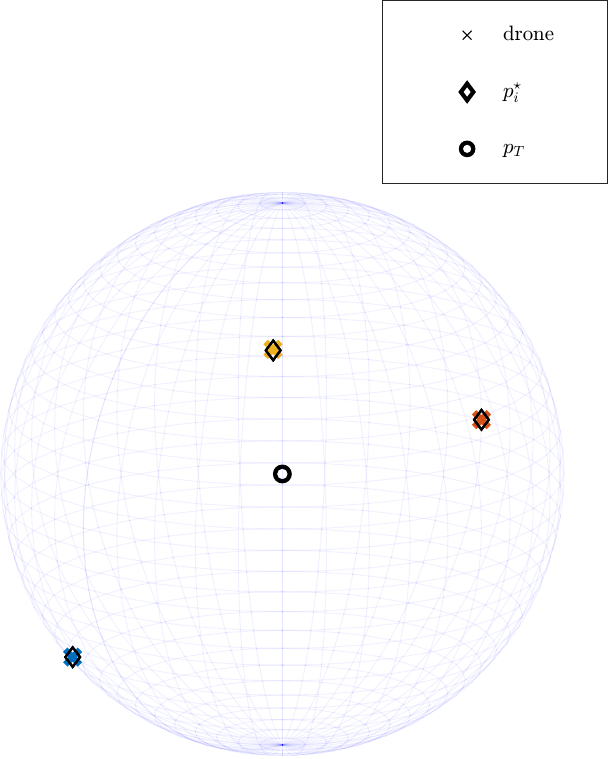}
    \\
    \includegraphics[width=0.45\columnwidth]{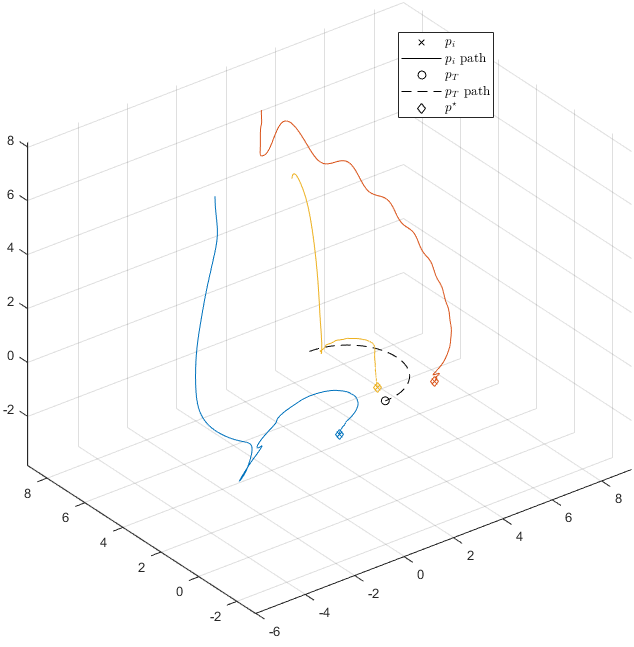}
    \includegraphics[width=0.45\columnwidth]{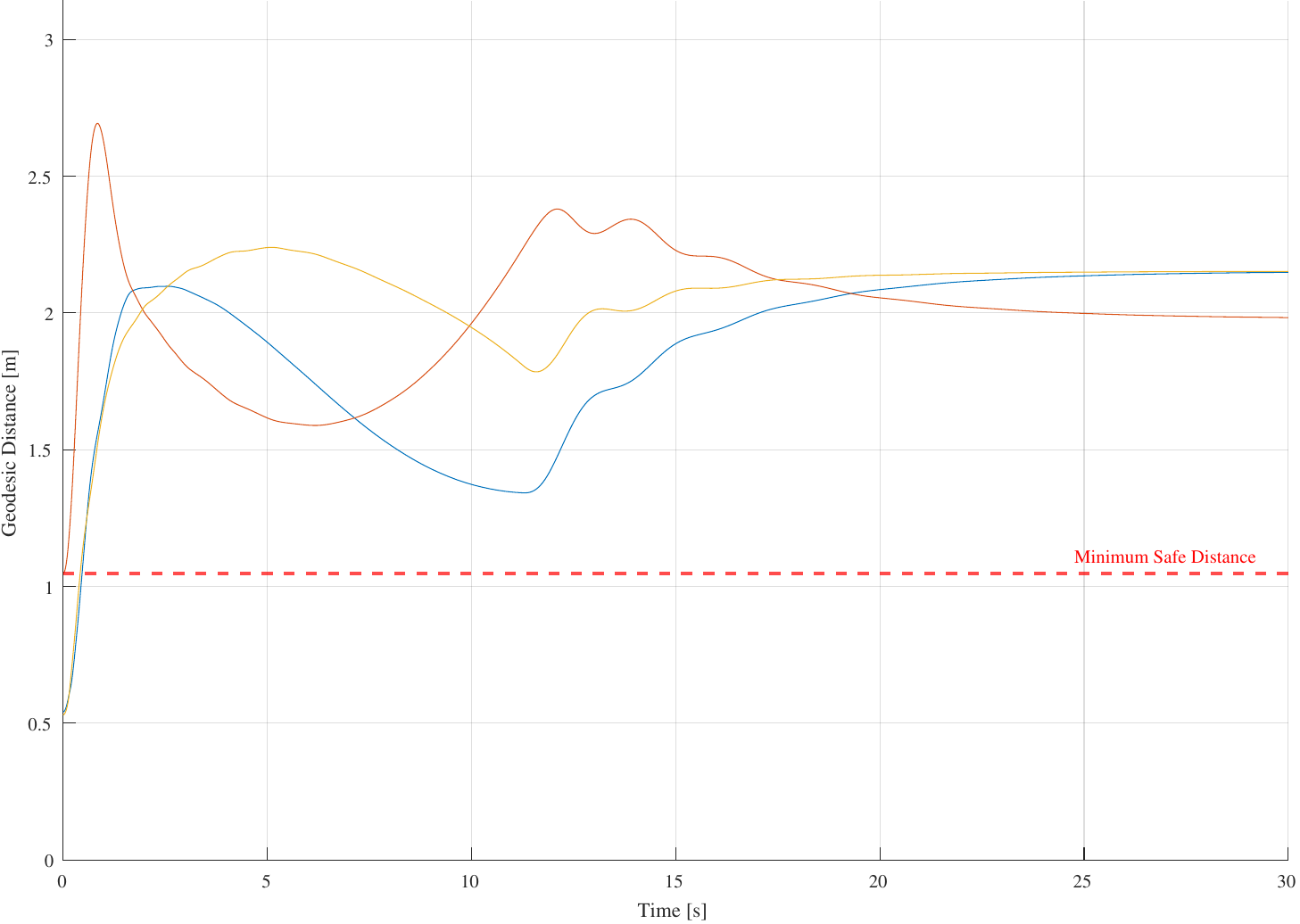}
    \caption{Simulation results for the 3-drone scenario.}
    \label{fig:results_3}
\end{figure}

\begin{figure}
    \includegraphics[width=0.45\columnwidth]{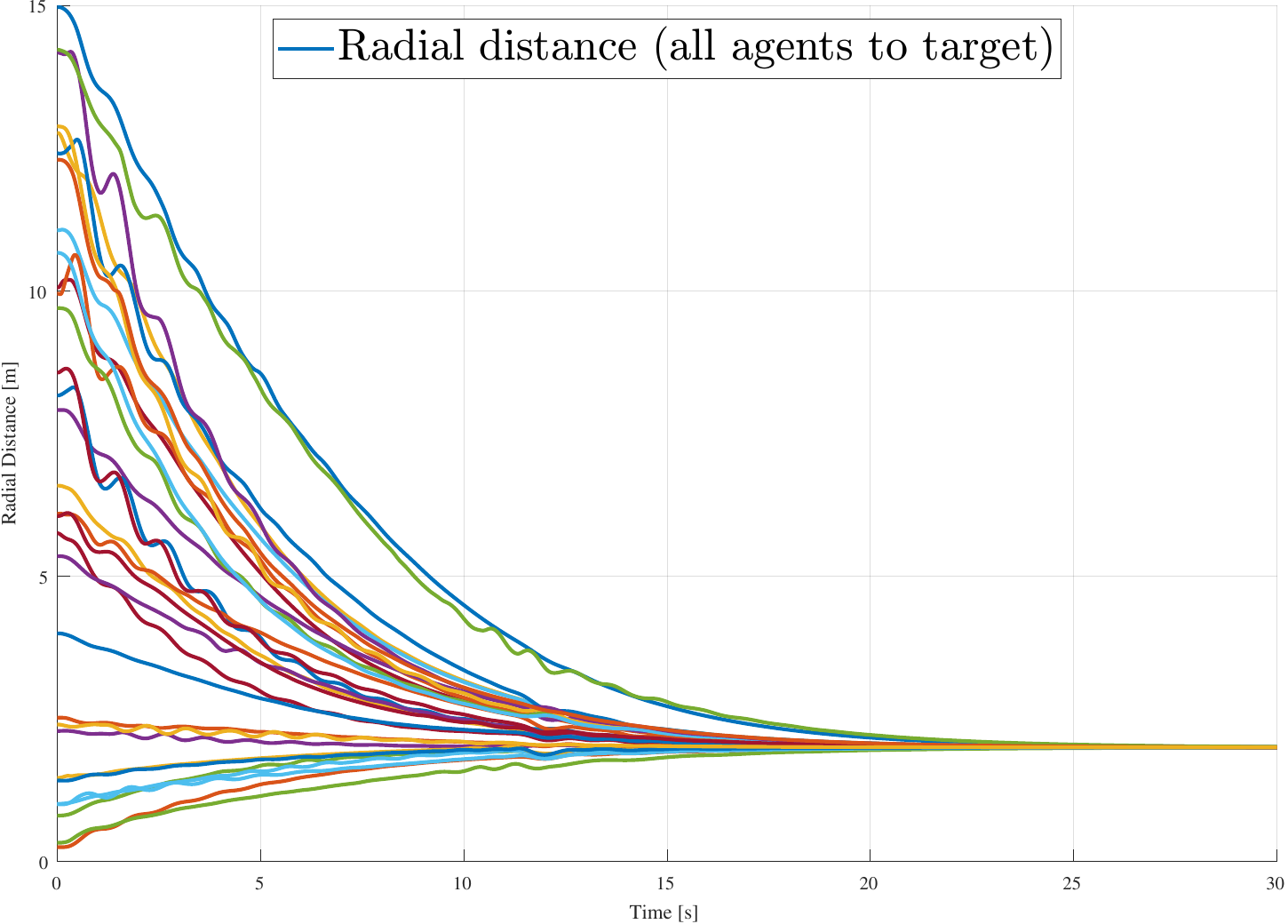}
    \includegraphics[width=0.45\columnwidth]{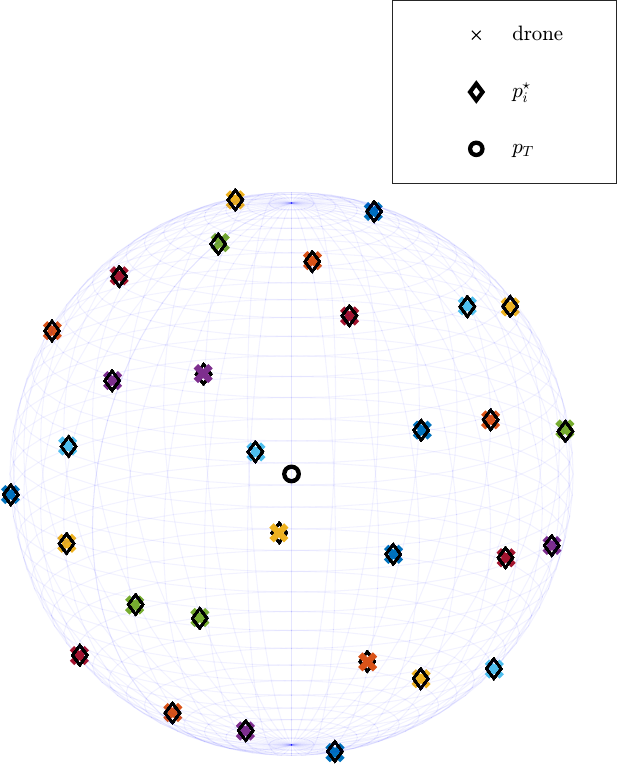}
    \\
    \includegraphics[width=0.45\columnwidth]{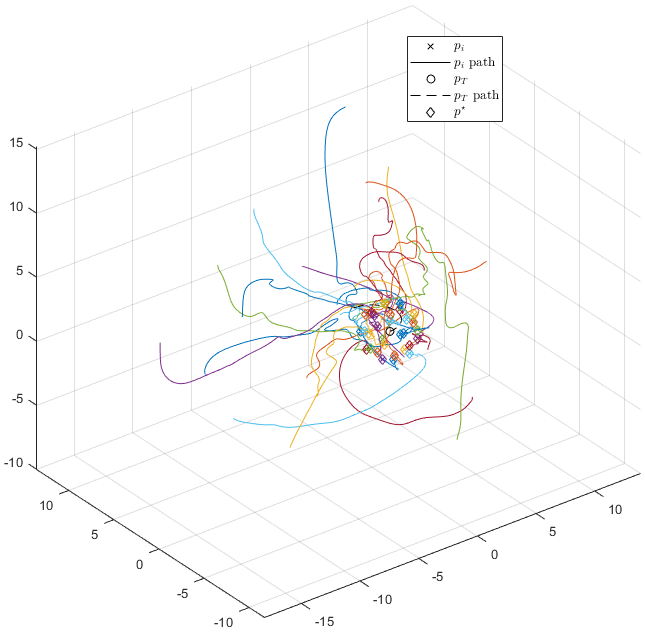}
    \includegraphics[width=0.45\columnwidth]{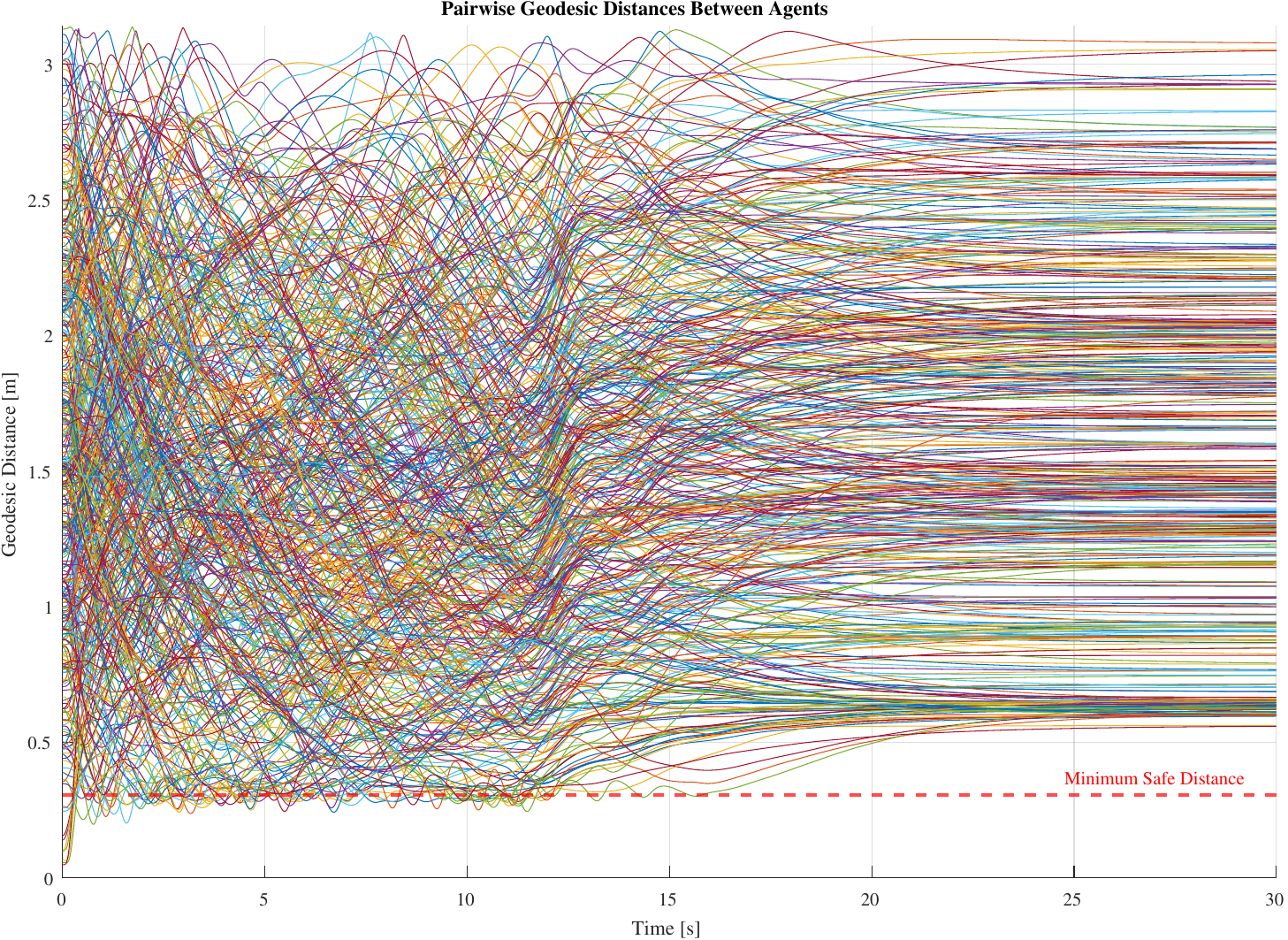}
    \caption{Simulation results for the 31-drone scenario.}
    \label{fig:results_31}
\end{figure}

Figure~\ref{fig:results_3} shows the results for the 3-drone scenario, including the reduction of the encirclement radius, final positions after 30 seconds, agent trajectories, and the evolution of inter-agent distances. Equivalent results for the 31-drone scenario are shown in Figure~\ref{fig:results_31}.

Both scenarios successfully achieve stable encirclement at the desired radius while maintaining safe separation between agents during convergence, as all inter-agent distances remain above the predefined safety threshold.

In the 3-drone case, the formation converges to an equilateral triangle with angular spacing of approximately $\frac{2\pi}{3}$. In the 31-drone scenario, having 465 pairwise agent connections, the swarm maintains adequate spacing throughout the maneuver, with angular separations generally remaining above the safety threshold. From the same graph it is possible to analyze that the convergence time is approximately 20 seconds.

\begin{table}
    \centering
    \caption{Simulation and Control Parameters}
    \label{tab:all_params}
    \begin{tabular}{l c}
    \hline
    \textbf{Parameter} & \textbf{Value} \\
    \hline
    \multicolumn{2}{c}{\textit{General Simulation Parameters}} \\
    \hline
    Simulation duration $T_{\mathrm{end}}$ & $30~\mathrm{s}$ \\
    Sampling period $\Delta t$ & $0.01~\mathrm{s}$ \\
    Gravity $g$ & $9.81~\mathrm{m/s^2}$ \\
    Number of drones $n_D$ & $3/31$ \\
    Desired sphere radius $r^\star$ & $2~\mathrm{m}$ \\
    Geometric gain $k_{\mathrm{geo}}$ & $5$ \\
    Radial gain $k_{\mathrm{rad}}$ & $0.3$ \\
    \hline
    \multicolumn{2}{c}{\textit{Nonlinear Controller Parameters}} \\
    \hline
    Mass $m_D$ & $5~\mathrm{kg}$ \\
    Inertia matrix $I_D$ & $\mathrm{diag}(0.02,\;0.02,\;0.03)$ \\
    Drag coefficient $D$ & $0$ \\
    Proportional gain $K_p$ & $\mathrm{diag}(150,\;150,\;150)$ \\
    Velocity gain $K_v$ & $\mathrm{diag}(15,\;15,\;15)$ \\
    Integral gain $K_i$ & $\mathbf{0}_{3\times3}$ \\
    Attitude gain $k_r$ & $\mathrm{diag}(8,\;8,\;8)$ \\
    Angular rate gain $K_\Omega$ & $\mathrm{diag}(0.5,\;0.5,\;0.5)$ \\
    \hline
    \end{tabular}
\end{table}

\section{Conclusions}\label{sec:conclusions}
A geometric framework for three-dimensional UAV encirclement based on the two-sphere has been presented. By combining point separation, Lyapunov-based reduced-attitude control and radial control the proposed approach guarantees almost global stable convergence. Proposed future work relies on the use of Control Barrier Function (CBF) either on the two-sphere $\Gamma^2$ or directly in $\mathbb{R}^3$ guaranteeing collision avoidance while evolving to the desired target.

\bibliography{refs}

\begin{thebibliography}{10}
\providecommand{\url}[1]{#1}
\csname url@samestyle\endcsname
\providecommand{\newblock}{\relax}
\providecommand{\bibinfo}[2]{#2}
\providecommand{\BIBentrySTDinterwordspacing}{\spaceskip=0pt\relax}
\providecommand{\BIBentryALTinterwordstretchfactor}{4}
\providecommand{\BIBentryALTinterwordspacing}{\spaceskip=\fontdimen2\font plus
\BIBentryALTinterwordstretchfactor\fontdimen3\font minus \fontdimen4\font\relax}
\providecommand{\BIBforeignlanguage}[2]{{%
\expandafter\ifx\csname l@#1\endcsname\relax
\typeout{** WARNING: IEEEtran.bst: No hyphenation pattern has been}%
\typeout{** loaded for the language `#1'. Using the pattern for}%
\typeout{** the default language instead.}%
\else
\language=\csname l@#1\endcsname
\fi
#2}}
\providecommand{\BIBdecl}{\relax}
\BIBdecl

\bibitem{jinCircularFormationControl2018}
L.~Jin, S.~Yu, and D.~Ren, ``Circular {{Formation Control}} of {{Multiagent Systems}} with {{Any Preset Phase Arrangement}},'' \emph{Journal of Control Science and Engineering}, vol. 2018, 2018.

\bibitem{kimCooperativeControlTargetcapturing2007}
\BIBentryALTinterwordspacing
T.~H. Kim and T.~Sugie, ``Cooperative control for target-capturing task based on a cyclic pursuit strategy,'' \emph{Automatica}, vol.~43, no.~8, pp. 1426--1431, Aug. 2007. [Online]. Available: \url{https://doi.org/10.1016/j.automatica.2007.01.018}
\BIBentrySTDinterwordspacing

\bibitem{zhengDistributedControlUniform2015}
R.~Zheng, Z.~Lin, M.~Fu, and D.~Sun, ``Distributed control for uniform circumnavigation of ring-coupled unicycles,'' \emph{Automatica}, vol.~53, pp. 23--29, 2015.

\bibitem{Cavalheiro2026Code}
F.~Cavalheiro, ``Hierarchical nonlinear control for swarm encirclement around a moving target uav,'' \url{https://github.com/Filipe-Cavalheiro/Hierarchical-Nonlinear-Control-for-Swarm-Encirclement-Around-a-Moving-Target-UAV}, 2026, gitHub repository. Accessed: 2026-07-27.

\bibitem{zhouFixedTimeTargetTracking2026}
Z.~Zhou, J.~Hu, X.~Wu, S.~Liao, and J.~Yuan, ``Fixed-{{Time Target Tracking}} and {{Encirclement Control}} for {{Multi-UAVs}} with {{Bearing-Only Measurements}},'' \emph{Drones 2026}, vol.~10, no.~1, p.~63.

\bibitem{litimeinSurveyTechniquesCircular2021}
\BIBentryALTinterwordspacing
H.~Litimein, Z.~Y. Huang, and A.~Hamza, ``A survey on techniques in the circular formation of multi-agent systems,'' \emph{Electronics (Switzerland)}, vol.~10, no.~23, Dec. 2021. [Online]. Available: \url{https://doi.org/10.3390/electronics10232959}
\BIBentrySTDinterwordspacing

\bibitem{wangMultiAgentBifurcationConsensusBased2023}
Y.~Wang, J.~Cao, and A.~Kashkynbayev, ``Multi-{{Agent Bifurcation Consensus-Based Multi-Layer UAVs Formation Keeping Control}} and {{Its Visual Simulation}},'' \emph{IEEE Transactions on Circuits and Systems I: Regular Papers}, vol.~70, no.~8, pp. 3221--3233.

\bibitem{patra3DCooperativePursuit2025}
D.~Patra and A.~R. Hota, ``{{3D Cooperative Pursuit}} with {{Guaranteed Enclosure}} via {{Robust Model Predictive Control}},'' in \emph{2025 {{American Control Conference}} ({{ACC}})}, 2025, pp. 2--7.

\bibitem{gonzalezMeasurementAreasSphere2010}
{\'A}.~Gonz{\'a}lez, ``Measurement of {{Areas}} on a {{Sphere Using Fibonacci}} and {{Latitude-Longitude Lattices}},'' \emph{Mathematical Geosciences}, vol.~42, no.~1, pp. 49--64, Jan. 2010.

\bibitem{mellingerMinimumSnapTrajectory2011}
\BIBentryALTinterwordspacing
D.~Mellinger and V.~Kumar, ``Minimum snap trajectory generation and control for quadrotors,'' in \emph{2011 {{IEEE International Conference}} on {{Robotics}} and {{Automation}}}, 2011, pp. 2520--2525. [Online]. Available: \url{https://doi.org/10.1109/ICRA.2011.5980409}
\BIBentrySTDinterwordspacing

\bibitem{rampAttitudeAngularVelocity2015}
M.~Ramp and E.~Papadopoulos, ``Attitude and angular velocity tracking for a rigid body using geometric methods on the two-sphere,'' in \emph{2015 {{European Control Conference}} ({{ECC}})}, pp. 3238--3243.

\end{thebibliography}

\end{document}